\documentclass[12pt,preprint]{aastex}

\usepackage{rotating}
\usepackage{multirow}
\usepackage{graphicx}

\newcommand{\monitor}{Monitor\footnote{http://www.solarmonitor.org/}}
\newcommand{\eisnotes}{\footnote{ftp://sohoftp.nascom.nasa.gov/solarsoft/hinode/eis/doc/eis\_notes/}}
\shorttitle{Emission measure in diffuse regions}
\shortauthors{Subramanian et al.}

\begin{document}

\title{Emission Measure Distribution for Diffuse Regions in Solar Active Regions}

 \author{Srividya Subramanian, Durgesh Tripathi}
 \affil{Inter-University Centre for Astronomy and Astrophysics, Post Bag-4, Ganeshkhind, Pune 411007, India}

 \author{James A. Klimchuk}
 \affil{NASA Goddard Space Flight Center, Greenbelt, MD 20771, USA}

\author{Helen E. Mason}
 \affil{Department of Applied Mathematics and Theoretical Physics, University of Cambridge, Wilberforce Road, Cambridge CB3 0WA, UK}

\begin{abstract}

Our knowledge of the diffuse emission that encompasses active regions is very limited. In the present paper we investigate two off-limb active
regions, namely AR10939 and AR10961, to probe the underlying heating mechanisms. For this purpose we have used spectral observations from
Hinode/EIS and employed the emission measure (EM) technique to obtain the thermal structure of these diffuse regions. Our results show that the
characteristic EM distributions of the diffuse emission regions peak at $\log\, T = 6.25$ and the cool-ward slopes are in the range 1.4 - 3.3. This
suggests that both low as well as high frequency nanoflare heating events are at work. Our results provide additional constraints on the
properties of these diffuse emission regions and their contribution to the background/foreground when active region cores are observed on-disk.

\end{abstract}


\section{Introduction} \label{intro}

Our knowledge on how the solar corona is heated and maintained at a million degrees
kelvin, when the photosphere below is at 5800 kelvin, is still far from being
comprehensive. Active Regions (ARs) are ideal observing targets to probe the underlying
heating mechanisms as they are the locations of profound heating processes. In
addition, they show a wide distribution of physical parameters. Accurate measurements
of such parameters are critical in the formulation and constraint of coronal heating theories.
For extended discussions on coronal heating, refer to, e.g., \cite{klimchuk06} and \cite{reale10}.

Topologically, ARs possess different structures like the core loops primarily seen at
$\approx$ 3-5 MK rooted in moss regions which are seen primarily at around 1-2 MK
\citep[and references therein]{tripathi10core, tripathi11core, viall12}, warm loops at $\approx$ 1 MK
\citep{giulio03,tripathi09,ignacio09}, the cool fan structures at the edges
of the ARs at $<$ 1 MK \citep{schrijver99,wine02,warren11,young12}. In addition to the visible loop structures,
there is a substantial amount of diffuse emission in and around the active regions \citep{giulio03,viall11,giulio14}.
Diffuse emission regions may be defined as regions with no resolvable structures. They may, however, appear
to be loop-like structures at higher resolution. In fact, the emissions from confined
loops structures are just about 20-30\% higher than the background/foreground diffuse
emission \citep{giulio03, viall11,giulio14}. \cite{odwyer11} investigated the density and temperature structure of a limb AR.
The authors reported that AR plasmas are multi-thermal and the
electron densities fall off as a function of distance from the core, as was also obtained using the
{white light} observations of the corona.

AR heating has often been debated between effectively steady heating (high frequency nanoflares)
and impulsive heating (low frequency nanoflares). In the former scenario, the delay between heating events is smaller than the cooling/draining timescale of the
plasma, leading to conditions that are similar to constant heating. Impulsive
heating, however, suggests that the delay between heating events is longer than
the cooling/draining timescale of the plasma, i.e. the plasma gets some time to cool
down before another heating event takes place. The properties of 1 MK warm loops seem
to favor impulsive heating \citep{warren03, ignacio09, tripathi09, klimchuk09}.
However, the heating of core loops is a matter of strong debate
\citep{tripathi10core, tripathi11core, wine11, warren11, warren12, viall12, viall13, dadashi12, bradshaw12, wine13}.
Recent analysis suggest that active regions
during {the early part} of their evolution seem to show an EM distribution {(EMD)} that is consistent with  impulsive heating,
while during the latter part of their evolution, the variability of the core becomes more gentle and the EM
distribution is more consistent with high-frequency nanoflare heating \citep{ignacio12,giulio14}.

The study of diffuse emission has not been explored in great detail. The main aim of this paper is to study and
probe the heating mechanism in the diffuse part
of active regions. Direct observations of heating processes are not possible yet, as these events
happen on scales much smaller than the resolvable limits with the available present day instrumentation
\citep{klimchuk06}. Emission measure (EM) diagnostics have been advocated \citep[see e.g.,][]{tripathi10core} to be one of the
possible {indirect} modes of studying the heating mechanisms among others such as Doppler shifts
\citep{brooks09, tripathi12, wine13, dadashi12} and the recently developed time lag analysis \citep{viall12}.

Here, we have employed the
technique of EM to study the diffuse emission in active regions. In addition, we estimate the EM(T)
distribution of topologically different regions in off limb \textit{AR 10939} and \textit{AR 10961}, i.e., warm and core loop
structures and compare with the EM(T) of the diffuse
emission regions with the aim to probe the thermal structure in diffuse regions and thereby their heating mechanisms. \cite{giulio13} presented a revised radiometric calibration, due to the degradation of detector's response of the EIS instrument over time, since the launch date of the Hinode mission. The revised calibration has been accepted by the EIS team and has been provided in the solarsoftware package. In the present work, we have applied the revised calibration on intensities to obtain the EM  and compared with that obtained using the unrevised intensities.
The rest of the paper is organised as follows: in section~\ref{obs}, we describe
the data used in this study and the reduction procedures applied; in section~\ref{db}, we discuss about an instrumental effect, the diffraction bands that are observed in our datasets; we describe our data analysis
method and present our results in section~\ref{analysis} followed by a summary and discussion in
section~\ref{conc}.

\section{Observations and Data reduction} \label{obs}

\begin{table*}[hp!]
\small
\caption{List of spectral lines used to study the EM over topologically different areas within active region 10939,
along with the calculated intensities and the line fitting errors, of a sample of masked regions (m2, m5, m8 \& m10) in the respective wavelengths. True errors will also include radiometric calibration errors of about 22\% of the intensities \citep{lang06}, in addition to the fitting errors. Refer to Table \ref{eis_linelist_AR10939_2} for the intensities of the rest of the masked regions. Intensity units are in ergs cm$^{-2}$ s$^{-1}$ sr$^{-1}$. The intensities given here are not corrected with the revised radiometric calibration presented by \cite{giulio13} and the required intensity corrections can also be found here.  }
\label{eis_linelist_AR10939_1}
\begin{tabular} {lcccrrrr}
\\
\hline
\multicolumn{4}{c}{Spectral lines} & \multicolumn{4}{c}{Intensities and fitting errors} \\
\hline
Ion & Wavelength & $\log\,T$&Intensity	& m2 & m5 & m8 & m10  	\\
    & 	($\AA$)  & (K)    &correction& 	& & & 	\\
    & 	  & 	& factors& 	& & &  	\\
\hline
\ion{Mg}{6}   	& 268.990  & 5.65 & 1.11  &   132 $\pm$  2 &  287 $\pm$  3 &  596 $\pm$  3 &  966 $\pm$  5 \\
\ion{Si}{7}  	& 275.350  & 5.80 & 1.14  &   110 $\pm$  1 &  215 $\pm$  2 &  426 $\pm$  2 &  631 $\pm$  4 \\
\ion{Fe}{8} 	& 185.210  & 5.80 & 1.35  &   336 $\pm$  4 &  626 $\pm$  6 & 1325 $\pm$  8 & 2278 $\pm$ 13 \\
\ion{Fe}{8}  	& 186.600  & 5.80 & 1.39  &   215 $\pm$  3 &  404 $\pm$  4 &  895 $\pm$  5 & 1352 $\pm$  9 \\
\ion{Fe}{9}    	& 197.860  & 5.90 & 1.16  &   128 $\pm$  1 &  176 $\pm$  1 &  251 $\pm$  1 &  247 $\pm$  2 \\
\ion{Fe}{9}    	& 188.500  & 5.90 & 1.43  &   195 $\pm$  2 &  271 $\pm$  3 &  400 $\pm$  3 &  438 $\pm$  4 \\
\ion{Fe}{10}   	& 174.531  & 6.05 & 1.57  &  1852 $\pm$ 82 & 2193 $\pm$ 97 & 2785 $\pm$ 96 & 2746 $\pm$123 \\
\ion{Fe}{10}   	& 184.537  & 6.05 & 1.35  &   794 $\pm$  6 &  910 $\pm$  7 & 1154 $\pm$  7 & 1265 $\pm$ 10 \\
\ion{Fe}{11}   	& 188.217  & 6.15 & 1.45  &  2393 $\pm$ 20 & 2664 $\pm$ 25 & 3207 $\pm$ 25 & 3673 $\pm$ 35 \\
\ion{Fe}{11}   	& 180.401  & 6.15 & 1.47  &  1513 $\pm$  6 & 1618 $\pm$  7 & 1927 $\pm$  8 & 2406 $\pm$ 11 \\
\ion{Si}{10}	& 258.374  & 6.15 & 1.35  &   882 $\pm$  5 & 1033 $\pm$  5 & 1292 $\pm$  5 & 2334 $\pm$  9 \\
\ion{Si}{10}	& 261.060  & 6.15 & 1.29  &   406 $\pm$  3 &  456 $\pm$  3 &  569 $\pm$  3 &  865 $\pm$  5 \\
\ion{Fe}{12}  	& 195.120  & 6.20 & 1.00  &  1093 $\pm$  3 & 1310 $\pm$  4 & 1558 $\pm$  4 & 2054 $\pm$  6 \\
\ion{Fe}{12}  	& 192.394  & 6.20 & 1.13  &  2781 $\pm$  4 & 3179 $\pm$  6 & 3727 $\pm$  6 & 5227 $\pm$ 10 \\
\ion{Fe}{13}    & 202.040  & 6.25 & 1.00  &  2757 $\pm$  9 & 3216 $\pm$ 10 & 3608 $\pm$  9 & 4147 $\pm$ 13 \\
\ion{Fe}{13} 	& 203.827  & 6.25 & 1.02  &  1712 $\pm$  9 & 2595 $\pm$ 10 & 3623 $\pm$ 10 & 9613 $\pm$ 20 \\
\ion{Fe}{14}   	& 274.204  & 6.30 & 1.10  &  1629 $\pm$  5 & 2300 $\pm$  6 & 3293 $\pm$  6 & 6290 $\pm$ 11 \\
\ion{Fe}{15}   	& 284.163  & 6.35 & 1.32  &  6527 $\pm$ 16 &11322 $\pm$ 20 &21790 $\pm$ 25 &39060 $\pm$ 44 \\
\ion{S}{13}	& 256.700  & 6.40 & 1.42  &   383 $\pm$  4 &  622 $\pm$  5 & 1444 $\pm$  7 & 3234 $\pm$ 13 \\
\ion{Fe}{16}  	& 262.980  & 6.45 & 1.25  &   344 $\pm$  3 &  779 $\pm$  4 & 2539 $\pm$  6 & 4337 $\pm$ 10  \\
\ion{Ca}{14}	& 193.870  & 6.55 & 1.04  &    18 $\pm$  1 &   94 $\pm$  1 &  435 $\pm$  2 &  541 $\pm$  3 \\
\ion{Ca}{15}	& 201.000  & 6.65 & 0.99  &    84 $\pm$ 10 &   61 $\pm$  3 &  280 $\pm$  3 &  313 $\pm$  5 \\

\hline
\end{tabular}
\end{table*}

\begin{table}[ht!]
\small
\centering
\caption{ List of spectral lines used to study the EM over topologically different areas within active region 10961,
along with the calculated intensities and the line fitting errors, of a sample of masked regions (m3, m7, m9 \& m11) in the respective wavelength.  True errors will also include radiometric calibration errors of about 22\% of the intensities \citep{lang06}, in addition to the fitting errors metioned in this table. Refer to Table~ \ref{eis_linelist_AR10961_2} for the intensities of the rest of the masked regions. Intensity units are in ergs cm$^{-2}$ s$^{-1}$ sr$^{-1}$. Note: The intensities given here are not corrected with the revised radiometric calibration presented by \cite{giulio13}. }
\label{eis_linelist_AR10961_1}
\begin{tabular} {lcccrrrr} \\ 
\hline
\multicolumn{4}{c}{Spectral lines} & \multicolumn{4}{c}{Intensities and fitting errors} \\
\hline
Ion & Wavelength & $\log\,T$&Intensity	 & m3 & m7 & m9 & m11 \\
    & 	($\AA$)  & (K)	&correction & & & & 	\\
    & 	  & 	& factors& 	& & & \\ 	
\hline
\ion{Si}{7} & 275.350  & 5.80 & 1.23  &   19 $\pm$  1 &   37 $\pm$  1 &  136 $\pm$  2 &  132 $\pm$  2 \\
\ion{Fe}{8} & 185.210  & 5.80 & 1.35  &   43 $\pm$  2 &  104 $\pm$  4 &  341 $\pm$  5 &  578 $\pm$  7 \\
\ion{Fe}{10}& 174.531  & 6.05 & 1.57  &  939 $\pm$ 84 & 1171 $\pm$122 & 1569 $\pm$127 & 1988 $\pm$153 \\
\ion{Fe}{10}& 184.537  & 6.05 & 1.35  &  275 $\pm$  3 &  351 $\pm$  5 &  591 $\pm$  6 &  726 $\pm$  7 \\
\ion{Fe}{11}& 188.217  & 6.15 & 1.45  &  682 $\pm$  3 &  781 $\pm$  5 & 1190 $\pm$  6 & 1614 $\pm$  8 \\
\ion{Fe}{12}& 195.120  & 6.20 & 1.00  & 1471 $\pm$  3 & 1724 $\pm$  4 & 2367 $\pm$  5 & 3289 $\pm$  7 \\
\ion{Fe}{13}& 202.040  & 6.25 & 1.00  & 1522 $\pm$  5 & 1883 $\pm$  8 & 2331 $\pm$  9 & 2889 $\pm$ 11 \\
\ion{Fe}{13}& 203.827  & 6.25 & 1.02  &  574 $\pm$  4 &  920 $\pm$  7 & 1739 $\pm$  9 & 3992 $\pm$ 15 \\
\ion{Fe}{14}& 274.204  & 6.30 & 1.18  &  604 $\pm$  3 &  953 $\pm$  5 & 1383 $\pm$  6 & 2555 $\pm$  9 \\
\ion{Fe}{15}& 284.163  & 6.35 & 1.42  & 1689 $\pm$  6 & 3240 $\pm$ 11 & 5258 $\pm$ 14 &10343 $\pm$ 22 \\
\ion{Fe}{16}& 262.980  & 6.45 & 1.34  &   57 $\pm$  2 &  164 $\pm$  3 &  409 $\pm$  4 &  953 $\pm$  6 \\

\hline
\end{tabular}
\end{table}

For the present analysis, we have used the off-limb \textit{AR 10939} and \textit{AR 10961} datasets obtained on
2007~Jan~26 and 2007~July ~08, respectively, from the Extreme ultraviolet Imaging Spectrometer \citep[EIS;][]{eis2,eis1}
on board Hinode. EIS rastered the the \textit{AR 10939} with a 1\arcsec~ slit over an area of 128\arcsec~$\times$~128\arcsec, with 26.5~sec exposure at each position {and with a step size of 1\arcsec}. The observing study of \textit{AR 10939} was a full spectral observation. Similarly, \textit{AR 10961} was rastered with a 1\arcsec~ slit over an area of
256\arcsec~$\times$~256\arcsec, with 15~sec exposure at each position {and with a step size of 1\arcsec}. This observing sequence consisted of 17 spectral windows and out of which 11 spectral lines have been used in the current study. Tables~\ref{eis_linelist_AR10939_1} $\&$ ~\ref{eis_linelist_AR10961_1} show the list of the spectral lines used in this study, along with their wavelengths and peak formation temperatures. The peak formation temperatures were obtained from recent Chianti ionization equilibrium calculations
\citep[Chianti v7.1;][]{dere97, chiantiv7pt1}.

The data are reduced using the standard procedure \textit{eis\_prep.pro}\eisnotes  ~and are corrected for EIS slit, tilt and satellite orbital
variation. Using \textit{eis\_autofit.pro}\eisnotes, a single Gaussian line fit is applied to
the data to derive the intensity maps, except for the \ion{Mg}{6}~268.99~{\AA}, \ion{Fe}{9}~197.86~{\AA}, {\ion{Fe}{11}~188.22~{\AA}}, \ion{Fe}{12}~195.12~{\AA}, \ion{Fe}{13}~203.83~{\AA}, {\ion{Ca}{14}~193.87~{\AA}} and \ion{Ca}{15}~201.00~{\AA}~lines. In the latter cases, multiple Gaussian fitting is applied to subtract the blended line contributions. The \ion{He}{2}~256~{\AA}\ line is one of the
EIS core lines, with the lowest formation temperature available in the EIS spectral range.
However, especially in the case of active regions, the interpretation
of this line gets complicated by blends with \ion{Si}{10}~256.37~{\AA}, \ion{Fe}{12}~256.41~{\AA}
and \ion{Fe}{13}~256.42~{\AA} \citep{young07}. Hence, we do not include the \ion{He}{2}~256~{\AA}\ line in this
work. The rest of the available lines are very weak in the off-limb structures and hence
could not be used. The \ion{Fe}{8}~185.21~{\AA}~ ($\log\, T = 5.6$) line is blended with the \ion{Ni}{16} line which peaks in the AR core plasma and de-blending is essential to estimate the \ion{Fe}{8} plasma parameters. In  \textit{AR 10961} data, the respective \ion{Fe}{8} line could not be de-blended because of the non availability of other \ion{Ni}{16} line observations. While in the case of \textit{AR 10939}, de-blending is performed by estimating the contribution of \ion{Ni}{16} to the \ion{Fe}{8} line through observing the \ion{Ni}{16}~195.27~{\AA} line.

Disentangling the diffuse emission regions from the rest of the active region is a non-trivial task, which is very crucial for this work.
Hence it is essential to track the active region from the center of the Sun to the limb, which may give some ideas regarding the
evolution of different structures in the active region. The six images in Fig.~\ref{fig1} display the intensity images of 
\textit{AR 10961} obtained in the \ion{Fe}{15} line from 2007 Aug 01 to 2007 Aug 08, tracking the evolution of the active region from
disk center to the limb. The \ion{Fe}{15} spectral line, whose peak formation temperature at $\log\, T \approx 6.35$, is the
strongest \rm{Fe} line in the EIS/Hinode spectrum at AR conditions and the core structures that are primarily at few million degrees can be
well studied at this temperature regime. Thus by tracking the particular AR core at \ion{Fe}{15} temperatures, the diffuse regions
could be confidently constrained to be well outside the core structures.

The intensity corrections required for the revised radiometric calibration were obtainied using \textit{eis\_ltds.pro} \citep[available in the ssw libraries;][]{giulio13}, which provides a factor by which the revised intensities differ with the unrevised ones. The revised intensities have been obtained by applying these correction factors to the initial intensities obtained from the data. From here onwards, we will be calling the initial intensities and the respective EM distributions as unrevised intensities and unrevised EMDs. While the intensities corrected according to the revised radiometric calibration and the corresponding EM distributions will be referred as revised intensities and revised EMDs.

\section{Diffraction bands}\label{db}

\begin{figure}[ht!]
\begin{center}
\hspace{-1cm}
\includegraphics[trim=5cm 2.5cm 4cm 3cm, clip=true,angle=180,scale=0.5]{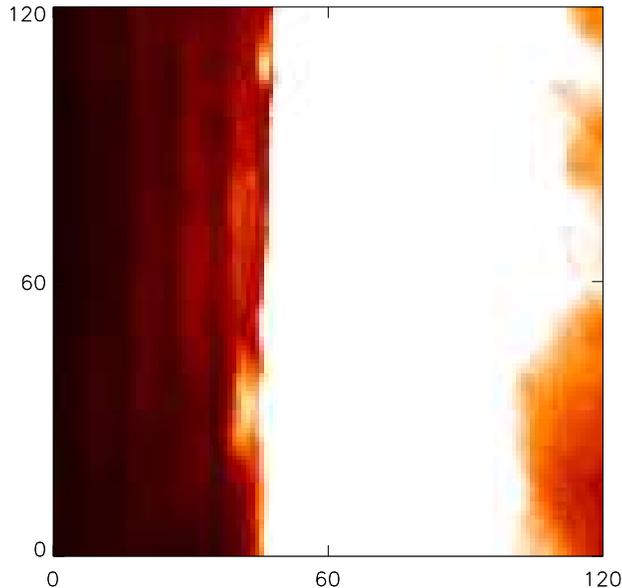}
\caption{Diffraction bands observed clearly on the \ion{Fe}{15}~intensity image of \textit{AR 10939}.}
\label{fig5}
\end{center}
\end{figure}

\begin{figure}[ht!]
\begin{center}
\hspace{-1cm}
\includegraphics[trim=2cm 1cm 3cm 2cm, clip=true,angle=180,scale=0.6]{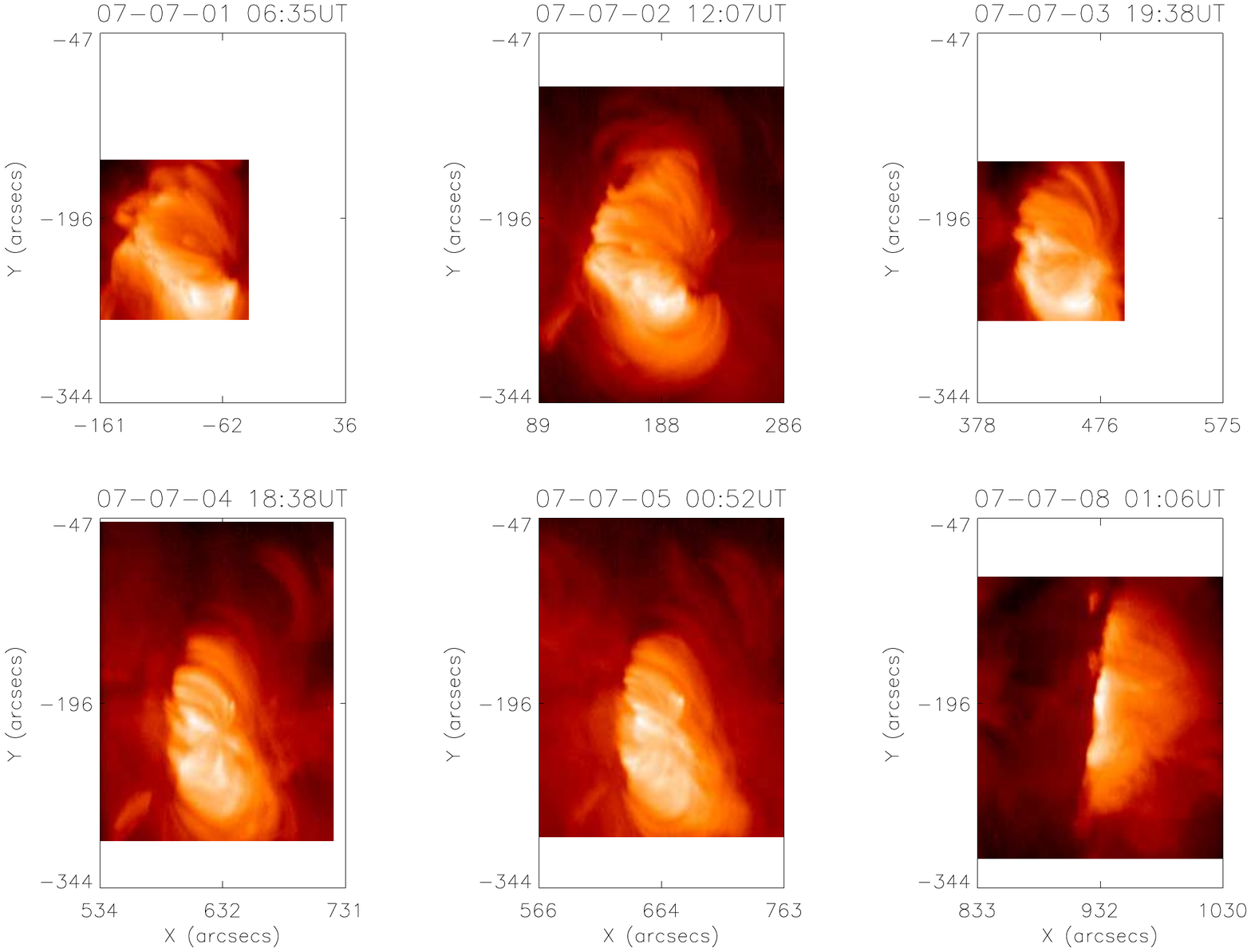}
\caption{\ion{Fe}{15}~intensity image of \textit{AR 10961} tracked from close to the center of the Sun (2007 Aug 01) to the limb (2007 Aug 08). All the images have been plotted with the same field of view of 297$\arcsec\times$197$\arcsec$, in order to help us in tracking any AR structure through the solar rotation. Though all the AR data have been plotted with the same y-axis scale, the scalings in the x-axis are different. This is because the apparant heliographic longitude (the solar X coordinate) of the AR changes vastly over the solar rotation, unlike the latitude (the solar Y coordinate).}
\label{fig1}
\end{center}
\end{figure}

The reduced data of both the above mentioned off-limb active regions show diffraction bands (Fig.~\ref{fig5}) in
the quiet regions beside the active regions. These bands are clearly seen in high temperature lines (\ion{Fe}{12} - \ion{Fe}{15}), being
most prominent in \ion{Fe}{15} line (Fig.\ref{fig5}). These are probably similar to the cross-shaped diffraction patterns seen in AIA
onboard SDO and TRACE during flares \citep[see e.g.,][]{raftery11, lin01}. In our cases, the scattered light forms bands rather than
the crossed-spikes patterns. This is probably due to the active region being a broad structure of enhanced brightness at the off-limb
when compared to the flaring regions (Young, P., 2013: private communication). The bands always trace the edges of the active regions and may arise due to the large intensity contrast produced by the active region over the off-limb background values.

{For \textit{AR 10939}, \ion{Fe}{15} intensity at the successive bands falls off approximately as 5\%, 3\%, 1\% and 0.3\% of the brightest part of the AR core ($\sim$ 39000~ ergs cm$^{-2}$ s$^{-1}$ sr$^{-1}$) at distances of approximately 10\arcsec, 20\arcsec, 30\arcsec, and 40\arcsec. Some of the observed emission at the locations of the bands comes from actual on-disk solar features, so these percentages represent upper limits on the diffracted core component. We use this information later to estimate the possible contamination in the diffuse emission above the limb.} These factors mentioned above would depend on how big/bright the source is in that particular raster. While the \textit{AR 10961} raster shows only a trace of such bands as the exposure time is much shorter (15~sec) than the exposure time of the \textit{AR 10939} raster (26.5~sec). \textit{AR 10961} data has another active region in the full field of view and it would be inappropriate to estimate the intensities of such bands as they 
have 
contamination from near-by structures. 

\section{Data Analysis and Results}\label{analysis}

The motivation behind this work is to probe the heating mechanism for the diffuse emission regions. Off-limb AR data are ideal to spectroscopically study the topologically different areas within the active region, especially the diffuse emission regions, as these regions could be isolated without much contamination from the core of the active regions as well as warm 1 MK loops. A detailed description on the spectroscopic techniques for deriving the physical parameters like electron density and temperature along with the EM of the emitting plasma has been given in {\citet{mason94} and \citet{tripathi10moss}}. {EMD} is a function of temperature and density of the emitting plasma. The observed intensity can be correlated with the emission measure as follows.

\begin{equation}
Intensity (I) = A_{el} \times  EM  \times C_{lamda}  \qquad \qquad \qquad \\
\end{equation}

\noindent where, $A_{el}$ is the elemental abundance and $C_{lamda}$ is the contribution function containing all the relevant atomic
parameters like transition probabilities, ionization fraction etc. Earlier works showed that, for ARs, EM obeys a power law $ EM(T)\propto T^\alpha $ up to a peak near 3 MK \citep{dere82, dere93}. While plotting the EM with respect to temperature on a log-log scale, $\alpha$ is the slope of a best fitted straight line up to the peak emission measure. The slope $\alpha$ represents the temperature distribution of multi-thermal plasma along the line-of-sight.

For both the ARs, \textit{eis\_pixel\_mask.pro} is used to choose a sample of topologically different areas within active regions
like hot core loops, comparatively cooler warm loops and the diffuse emission regions, by masking pixels in polygon mode. It
is crucial to make sure that each box contains only one type of feature, i.e., they should not have any contamination
from other regions. We have used \textit{eis\_mask\_spectrum.pro} to retrieve the spatially averaged spectra over each chosen
pixel group (box).

\begin{figure*}[htp!]abs(xx1[1:29])
\begin{center}
\includegraphics[trim=3cm 6cm 3cm 8cm, clip=true,angle=90,scale=0.5]{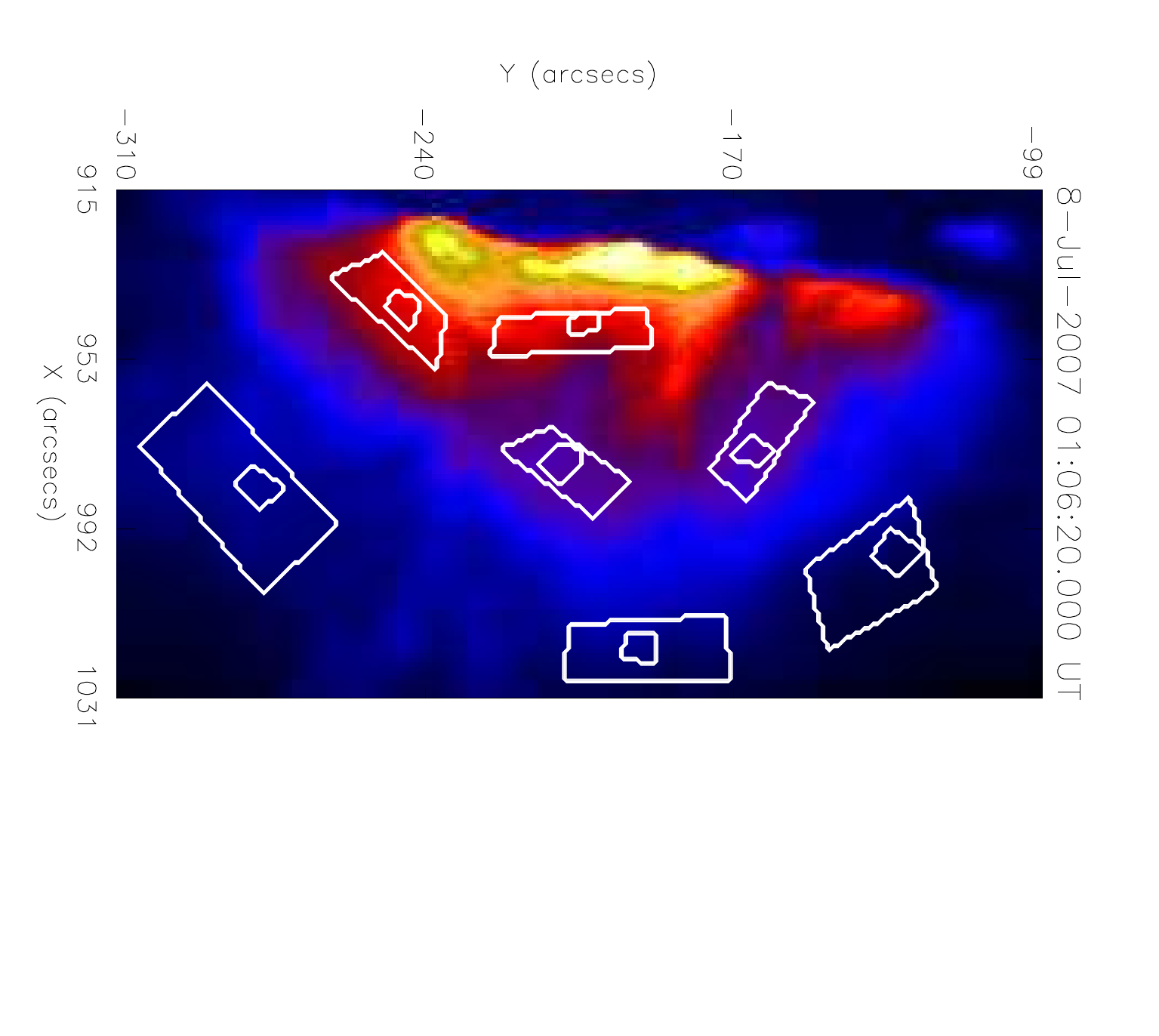}
\hspace{-1cm}
\includegraphics[trim=1cm 2cm 2cm 2cm, clip=true,angle=90,scale=0.4]{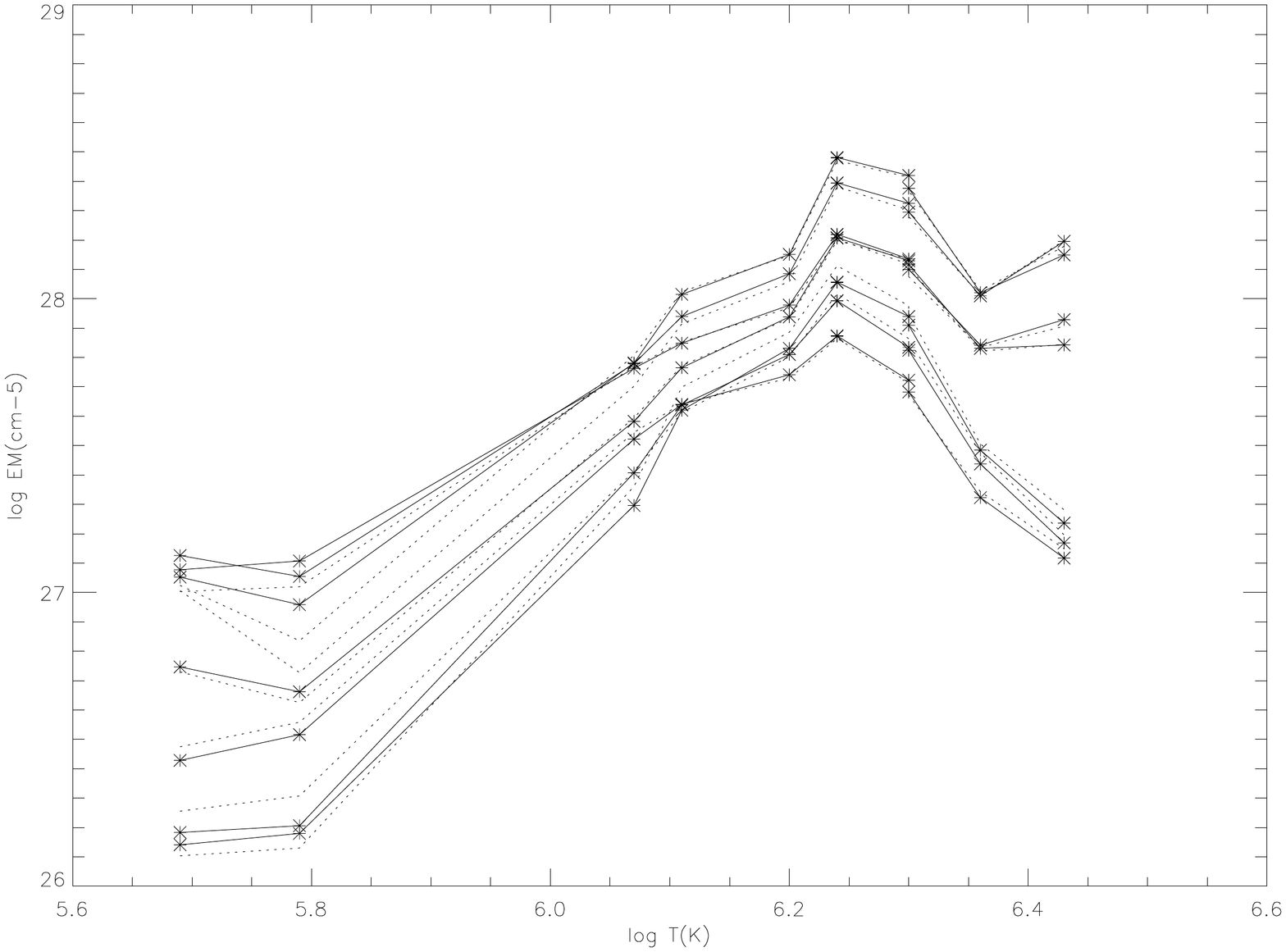}
\vspace{-0.5cm}
\caption{Left: Intensity image of AR 10961 in \ion{Fe}{13} with the chosen masked regions over-plotted with white boxes. Right: Emission measure curves of all the masked regions, where continuous and dotted lines represent masked regions enclosed by the bigger boxes and sub-boxes, respectively).   }
\label{fig2}
\end{center}
\end{figure*}

Estimation of emission measure from averaged spectra is non trivial. As the masked regions are selected manually, the box
sizes are not the same for all the regions. It is important to understand the effect of such varying box size in the
estimation of EM, in order to compare our results with the other similar works. Hence, we choose a set of seven randomly oriented
boxes each having a sub-box over \textit{AR 10961}. Thus, in total there are 14 regions as shown in Fig. \ref{fig2} (left) for the
analysis and their respective EM plots are shown in Fig. \ref{fig2} (right). The EM plot clearly shows that choosing different boxes with
different sizes, in a particular area in the field of view, does not vastly change the EM.

\begin{figure*}[htp!]
\begin{center}
\vspace{-1.5in}
\hspace{-0.2in}
\includegraphics[trim=1cm 5cm 2cm 1.5cm, clip=true,angle=180,scale=0.6]{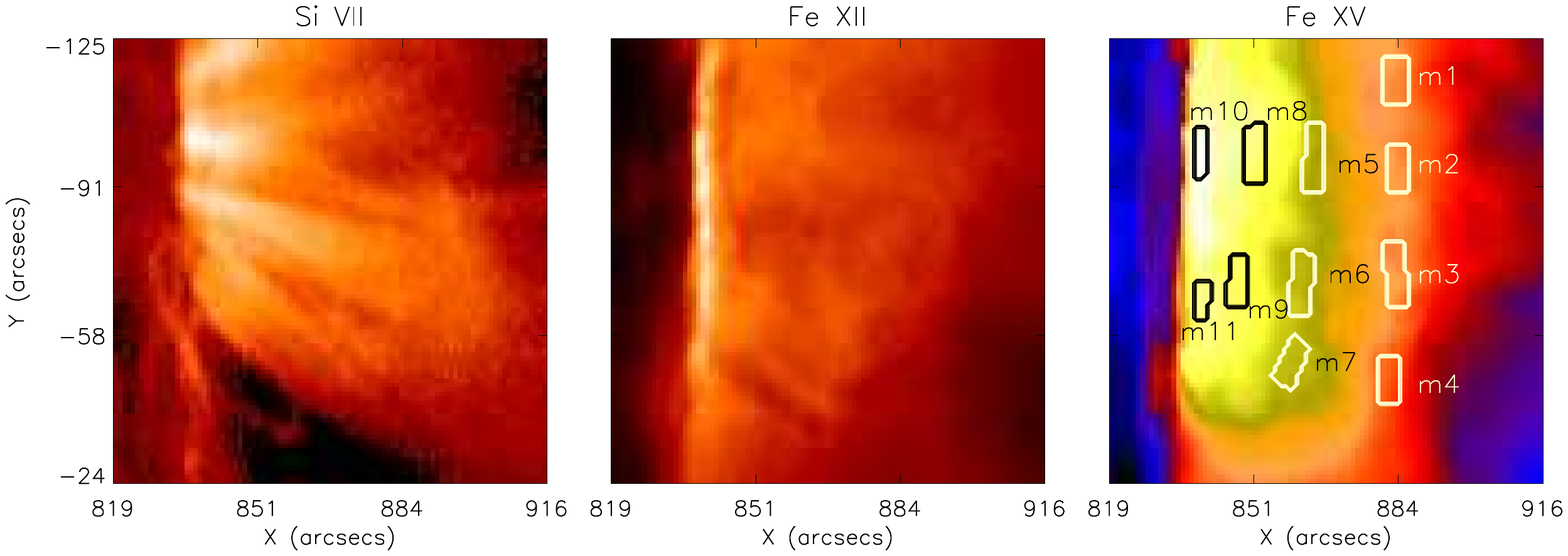}\\
\vspace{-0.1in}
\hspace{-0.2in}
\includegraphics[trim=0.4cm 0.5cm 0cm 0cm, clip=true,angle=90,scale=0.6]{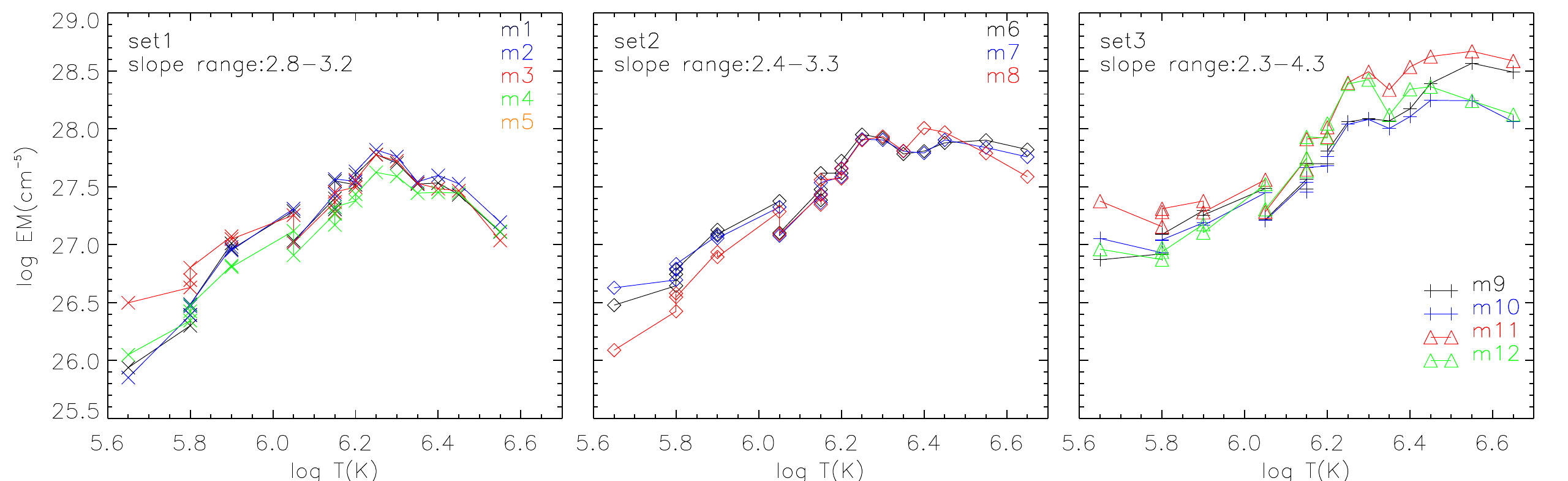}\\
\hspace{-0.2in}
\includegraphics[trim=0.4cm 0.5cm 0cm 0cm, clip=true,angle=90,scale=0.6]{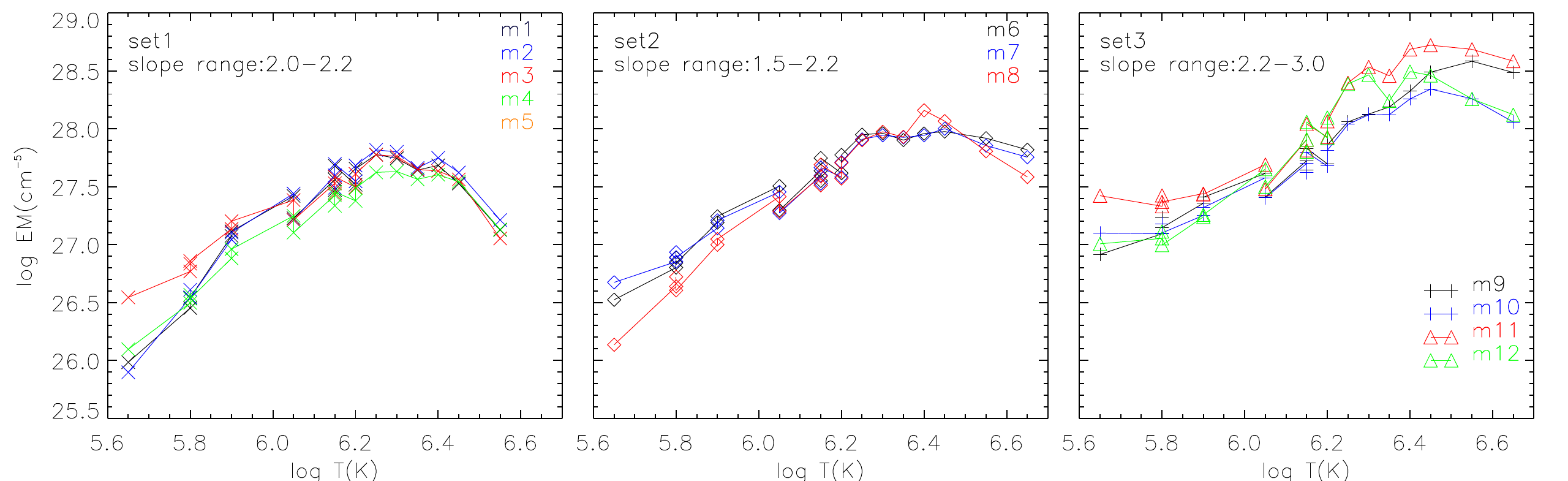}\\
\hspace{-0.2in}
\includegraphics[trim=0cm 0.5cm 0cm 0cm, clip=true,angle=90,scale=0.6]{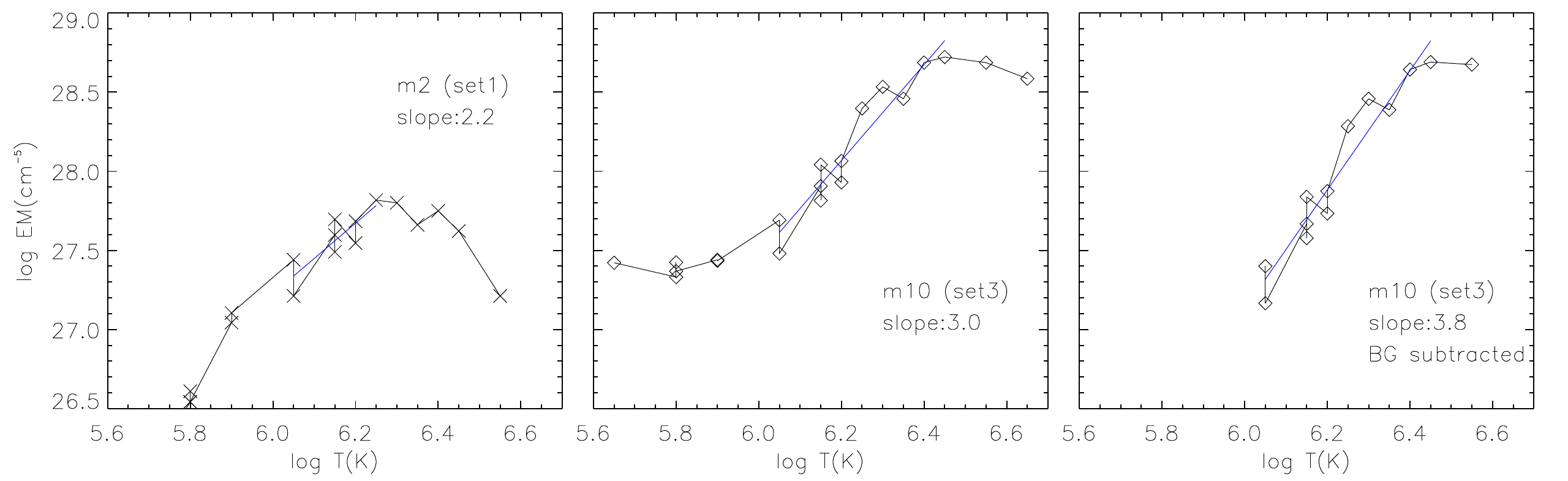}\\
\vspace{-0.5cm}
\caption{Top (row 1): Intensity image of \textit{AR 10939} in  \ion{Si}{7} (left), \ion{Fe}{13} (middle) and \ion{Fe}{15} (right) with white boxes (m1, m2, ..., m11) representing the 11 masked regions choosen for DEM analysis. Middle rows show the EMDs of all the masked regions estimated with unrevised (row 2) and revised (row 3) intensities, respectively.{The bottom row shows the revised EMDs of a sample diffuse region (left), a core region (middle) and the same core region after background subtraction (right), with their corresponding linear fits and the derived slopes.}}
\label{fig3}
\end{center}
\end{figure*}

\begin{figure*}[htp!]
\begin{center}
 \vspace{-1.5in}
\hspace{-0.2in}
\includegraphics[trim=0.5cm 5cm 2cm 2cm, clip=true,angle=180,scale=0.6]{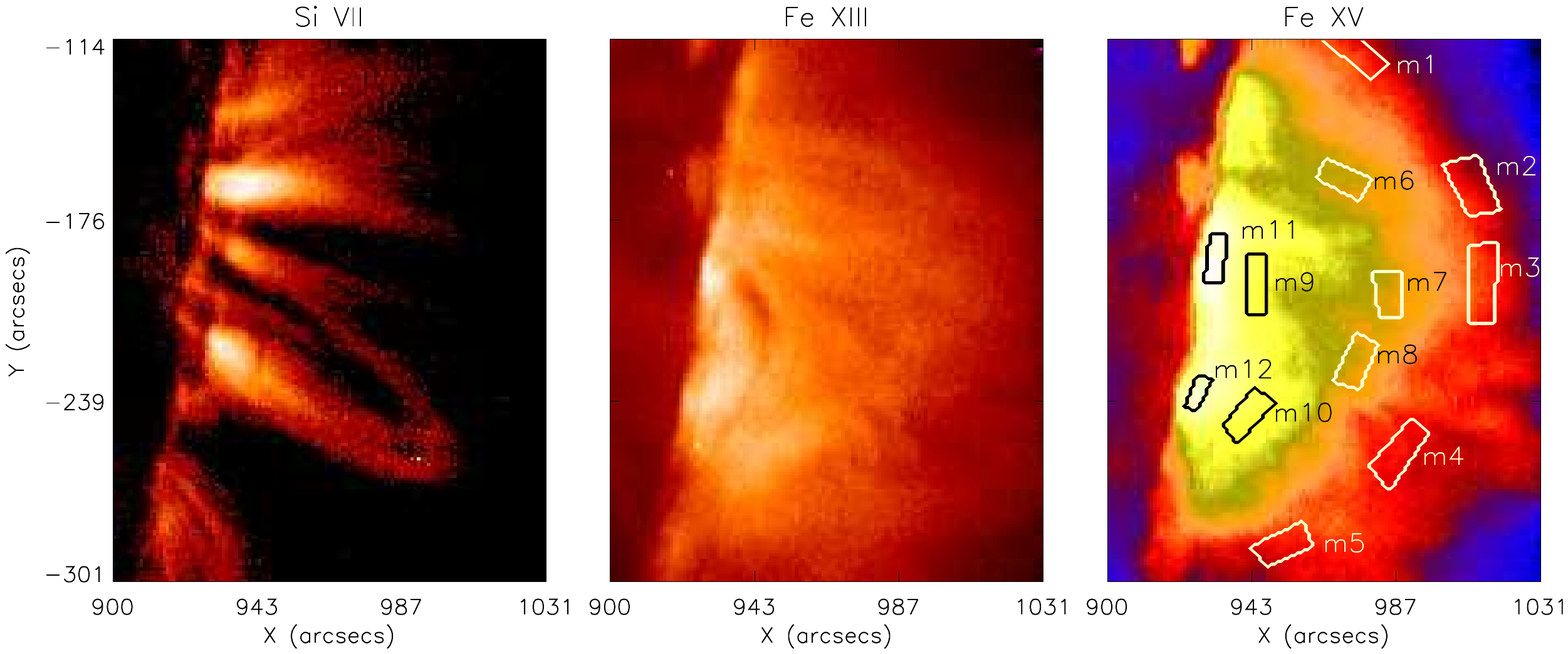}\\
\vspace{-0.1in}
\hspace{-0.2in}
\includegraphics[trim=0.3cm 0.5cm 0cm 0cm, clip=true,angle=90,scale=0.6]{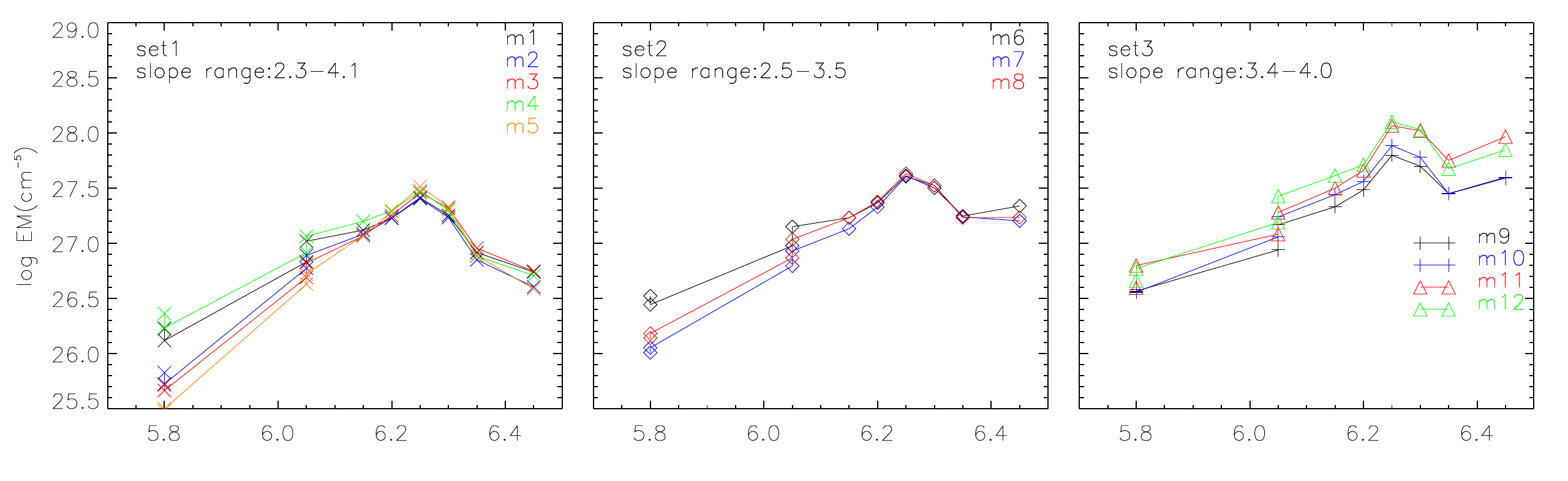}\\
\vspace{-0.1in}
\hspace{-0.2in}
\includegraphics[trim=0.2cm 0.5cm 0cm 0cm, clip=true,angle=90,scale=0.6]{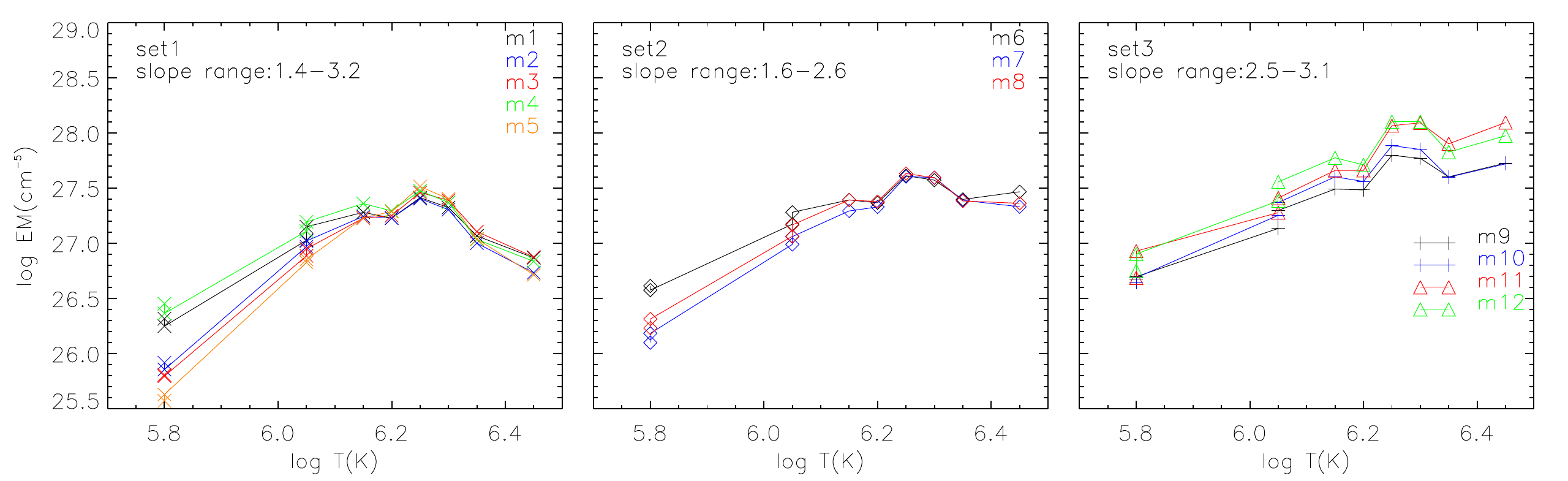}\\
\vspace{-0.3cm}
\caption{Top (row 1): Intensity image of the AR 10961 in  \ion{Si}{7} (left), \ion{Fe}{13} (middle) and \ion{Fe}{15} (right) with white boxes (m1, m2, ..., m12) representing the 12 masked regions choosen for EM analysis.  Bottom rows show the EM distributions of the 12 masked regions estimated with unrevised (row 2) and revised intensities (row 3), respectively. }
\label{fig4}
\end{center}
\end{figure*}

The color scheme (loadct, 5), which we have chosen for \ion{Fe}{15} images (Fig. \ref{fig2} (left) and  Fig. \ref{fig3} \& \ref{fig4}
(top right)), is found to be the best in representing the AR. The transition of color from blue to yellow distinctly represents the
topologically different regions in the AR from the diffuse parts on the outskirts of the AR towards the core, respectively.
Fig. \ref{fig3} (top panel) and \ref{fig4} (top panel) display the intensity images for \textit{AR 10939} and \textit{AR 10961},
respectively, obtained in \ion{Si}{7} (left), \ion{Fe}{13} (middle) and \ion{Fe}{15} (right) lines. \ion{Fe}{15} intensity image of
\textit{AR 10939} (Fig. \ref{fig3}: top right) are over-plotted with the 11 masked regions (m1, m2,  ..., m11) that are chosen for
the estimation of emission measure. We have divided these regions into different sets based on their locations. Set 1 is
comprised of the first four regions, namely m1, m2, m3 \& m4, which are selected in the diffuse emission region well above the
AR core. Set 2 is comprised of regions m5, m6, \& m7, which are located at the boundary between the core and the diffuse
emission region. Set 3 is comprised of regions m8, m9, m10 and m11 being in the core of the active region. Similarly, the
\ion{Fe}{15} intensity image of \textit{AR 10961} (Fig. \ref{fig4}: top right) is over-plotted with the selected 12 masked
regions (m1, m2, ..., m12). {These regions are again divided into four groups following the same criteria as discussed above in the case of \textit{AR~10939}.}

Initially, averaged spectra have been obtained from all the masked regions from \textit{AR~10939} and \textit{AR~10961} in all
the spectral windows mentioned in Table~\ref{eis_linelist_AR10939_1} and Table~\ref{eis_linelist_AR10961_1}, respectively.
Spectral lines are fitted with a modified version of \textit{eis\_autofit.pro} to read the output structures from
\textit{eis\_mask\_spectrum.pro}. For \textit{AR 10939}, estimated intensities for a sample masked region from each set
(m2 - set1, m5 - set2, m8 \& m10 - set3) are given in the Table~\ref{eis_linelist_AR10939_1} and the intensities of the rest of the
masked regions are given in Table~\ref{eis_linelist_AR10939_2}. Similarly for \textit{AR 10961}, estimated intensities for a sample
region from each set of masked regions (m3 - set1, m7 - set2, m9 \& m11 - set3) are given in the Table~\ref{eis_linelist_AR10961_1}
and the intensities of the rest are given in Table~\ref{eis_linelist_AR10961_2}. The masked regions discussed in the
Table~\ref{eis_linelist_AR10939_1} and the Table~\ref{eis_linelist_AR10961_1} fall as a  strip, with which variations of spectroscopic
parameters as a function of distance from the core of the AR can be explored. The EM is derived using the method discussed by
\citet{pottasch63, jordan71, tripathi10moss}. {Even though there are many true inversion DEM codes currently available, this method is objective and provides consistent results when compared to the other methods. Also, the results obtained using Pottasch method are similar to those from the other DEM inversion codes}. In order to derive the emission measure, it is necessary to provide electron density as an input parameter.  Therefore, we have used \ion{Fe}{13} (202.04~{\AA}~$\&~$203.83~{\AA}) lines to derive the densities for each region, as they are one of the best coronal density diagnostics available to EIS. Table \ref{table_dens} shows the estimated densities of all the masked regions choosen over both the ARs. For each spectral line, the EM at the peak formation temperature, $T_{max}$, is approximated by assuming that the contribution function is constant and equal to its average value over
the temperature range $T_{max}\pm$0.15 and zero at all other temperatures. We use the ionization equilibrium values as given in
Chianti \citep[v7.1;][]{dere97, chiantiv7pt1} and consider both photospheric \citep{grevesse98} and coronal \citep{feldman92} abundances.

\begin{table*}[ht!]
\small
\begin{center}
\caption{The electron densities obtained using \ion{Fe}{13}~202.040~{\AA} and 203.830~{\AA} of all the masked regions chosen over \textit{AR 10939}
and \textit{AR 10961}. Set1 represents the diffuse emission region, set3 represents the AR core regions and the set2 is the boundary between the AR core and the diffuse regions. }
\label{table_dens}
\begin{tabular} {rrrrrr}
\hline
\multicolumn{1}{c}{} & \multicolumn{2}{c}{\textit{AR 10939}} & \multicolumn{2}{c}{\textit{AR 10961}} \\
\hline
  & Masked & Density  & Masked& Density  	\\
  &  regions &(n$_e$~ cm$^{-3}$)&  regions&($n_e$~ cm$^{-3}$)	\\
\hline

set1&	m1	& 8.0e+08	&	m1	& 5.9e+08 \\
&	m2	& 8.6e+08	&	m2	& 5.3e+08 \\
&	m3	& 8.5e+08	&	m3	& 5.4e+08 \\
&	m4	& 7.1e+08	&	m4	& 5.7e+08 \\
&	{--}	&	{--}	&	m5	& 5.5e+08 \\
set2&	m5	& 1.2e+09	&	m6	& 7.8e+08 \\
&  	m6	& 1.1e+09	&	m7	& 7.1e+08 \\
&  	m7	& 1.2e+09	&	m8	& 7.5e+08 \\
set3& 	m8	& 1.5e+09	&	m9	& 1.1e+09 \\
& 	m9	& 1.5e+09	&	m10	& 1.3e+09 \\
&  	m10	& 6.5e+09	& 	m11	& 2.4e+09 \\
&	m11	& 4.1e+09	&	m12	& 2.4e+09 \\

\hline
\end{tabular}
\end{center}
\end{table*}

As listed in Tables~\ref{eis_linelist_AR10939_1},\ref{eis_linelist_AR10961_1},
\ref{eis_linelist_AR10939_2} and \ref{eis_linelist_AR10961_2}, the intensities in each spectral line show an increasing trend from the diffuse emission regions (set1) towards the core of the AR (set3). The \ion{Fe}{13} electron densities also show an increasing trend from set~1 to set~3. In each set, the derived electron densities are remarkably similar among the choosen masked
regions (Table \ref{table_dens}), except for the m10 $\&$ m11 (set3). The regions show higher densities when compared to m8 $\&$ m9 (set3) and also vary among themselves, which could be because of the possible contaminations from the underlying moss regions.
These trends observed in intensities and densities suggest that similar structures were
chosen in each set and also discriminate the regions chosen among different sets to be topologically different.

{The middle rows (row 2 and 3)} in Figure~\ref{fig3} display the EM distribution for all the different sets of masked regions chosen over \textit{AR 10939}, estimated with unrevised and revised intensities, respectively. In both the cases, the EM distribution for sets 2 and 3 are shown in the middle and right panels of rows 2 and 3. The EMDs belonging to set3 are plotted with two different symbols. The regions m8 and m9 are plotted with plusses and m10 and m11 are plotted with triangles. This is to make a distinction between core regions without and with possible moss regions contamination. Though almost all the curves of set2 and set3 show very similar trend, the EMDs for all the regions in set3 lie above those for set2. In addition, the peaks of the EM curves for all the regions in set3 are at larger temperatures than those of set2 regions. The peak of the emission measure for set3 mostly lies at around $\log\, T = 6.55$. However, the EM
distribution for regions m10 and m11 in set3 (plotted with triangles) also shows a definite peak at $\log\, T = 6.25$ which is characteristic of moss regions \citep{tripathi10moss}. In comparison, the EM distribution of set2 peaks at a lower temperature of $\log\, T\sim6.3$.  The revised radiometric calibration broadens the peak of {two (m5 $\&$ m6) out of the three masked regions choosen in this set, with the EM decreasing more slowly with temperature in the range $\log\, T = 6.25 - 6.45$. While, the third one (m7) which shows a double peak with the original calibration, namely at $\log\, T = 6.25$ and $\log\, T = 6.40$, shows only the second peak at $\log\, T = 6.40$ prominently with the revised calibration.}

We have studied the slopes of the EM and hot to warm ratios for the regions in set2 and set3. The ratio of hot to warm emission is determined by taking the ratio between the peak EM and the EM obtained at \ion{Fe}{10} and \ion{Si}{7}. The slope, however, is obtained between the temperature corresponding to peak EM and the EM of \ion{Fe}{10} at $\log\, T = 6.05$. The slope of unrevised EM for sets~2 \& 3 ranges between 2.3 $\&$ 4.2 and the ratios of hot to warm emission varies between 3.8 $\&$ 7.3 when warm emission is considered at $\log\, T = 6.05$. These ratios increase to the range between 9 $\&$ 28 when considering \ion{Si}{7} as warm emission. The revised calibration makes the slope comparatively shallower in the range of 1.5 $\&$ 3, with the hot to \ion{Fe}{10} emission varying between 2.8 $\&$ 5.4, and the hot to \ion{Si}{7} varying between 8 $\&$ 24.

The left panels of row 2 $\&$ 3 in Figure~\ref{fig3} show the emission measure distribution of diffuse regions classified as set1, comprising m1, m2, m3 \& m4. Both EM distribution curves (unrevised and revised) for all four regions are strikingly similar and show a definite peak at $\log\, T = 6.25$, i.e., at the peak formation temperature of \ion{Fe}{13}, except that the peak of revised EMDs is broader at \ion{Fe}{13} $\&$ \ion{Fe}{14} temperatures. For each of these curves, the estimated value of EMD slopes, $\alpha$, along with the emission measure ratio of the hot plasma to the warm plasma (\ion{Fe}{10} and  \ion{Si}{7}) are given in the Table \ref{table_slope_EMratio}. The slopes are computed between $\log\, T = 6.25$ and $\log\, T = 6.05$. The active region spectra for \textit{AR 10939} are full spectral scans, and the obtained EMD provide a complete picture of the thermal structure of the regions. Therefore, we conclude that the peak of the EM distribution for diffuse regions lies at $\log\, T=6.25$.

{The masked regions choosen in the sets 1, 2, and 3 (m8$\&$m9) are located approximately 50", 25", 10" from the brightest part of the core (where set3: m10$\&$m11 fall). Using the results of Section 3, we estimate a maximum contamination from diffracted core emission at these locations in \ion{Fe}{15} to be less than 0.3\%, 2\%, and 5\% respectively. This is about 117, 780 and 1950~ ergs cm$^{-2}$ s$^{-1}$ sr$^{-1}$, which is less than 2\%, 7\%, and 9\% of the actual observed \ion{Fe}{15} intensity at these locations. Therefore, it is safe to conclude that the contamination from the diffracted emission does not significantly impact our EM measurements.

If one assumes that the diffuse emission seen above the core in set1 is equivalent to the diffuse emission present along the line of sight in the studied core regions in set3, i.e., equivalent to the foreground/background emission along the line of sight, then it is possible to subtract the diffuse contribution from set3 to obtain the emission from the core plasma itself. When we do this using an averaged EMD from all the masked regions in set1, the EM slopes of the core regions (set3) increase by {0.64 to  0.94}, i.e., the distributions become steeper.  {The bottom row (row 4) in Figure~\ref{fig3} shows the best linear fits to a sample diffuse region (left), a core region (middle) and the same core region after background subtraction (right), with the corresponding slopes. } The core observations may also be contaminated by warm loops in the foreground and background, so this must be considered only a partial correction. Because of their characteristic EMD peaking at \ion{Fe}{13}, slope corrections using 
diffuse region EMD are expected to make the core region distribution steeper.} We caution, however, that the magnitude of the correction is very uncertain. Both the emissivity and the line-of-sight depth of the diffuse plasma could be considerably different at the altitude of set3 than at the altitude of set1. 

AR core EMDs (set 3) show a clear rollover at about $\log\, T = 6.3$ and this is most likely due to the contamination from the underlying moss regions, whose EMDs are known to peak at $\log\, T = 6.25$ \citep{tripathi10moss}. The EMD slopes would get steeper, by about a factor 0.2 (set 3a) and 0.8 (set 3b), when estimated with respect to this rollover peak, rather than with respect to the EM peak. Since we are interested primarily in the coronal loop top emission and not the emission from footpoint moss regions, the slopes presented here are estimated with respect to the peak EM.

{\cite{tripathi11core} studied \textit{AR 10961} when it was near disk center, when full spectral scan observations were available. Their goal was to measure the EM slope of inter-moss regions (i.e., of the AR core). Because of the on-disk observing geometry, lines-of-sight through the inter-moss regions included contributions from higher altitude diffuse emission above the core. To estimate to the magnitude of these contributions, they examined additional lines-of-sight outside the core and made adjustments to account for gravitational stratification. The diffuse emission was thereby estimated to contribute between 5 and 40$\%$ of the total EM at $\log\, T = 5.8$ along the line of sight through the core. Our limb measurements of this same active region give compatible results. Fig.~\ref{fig4} shows that, at this temperature, the diffuse emission (set1) is roughly 20$\%$ of the core emission (set3). The boundary region emission (set2) is roughly 30$\%$ of the core emission. This suggests that the method used 
by \cite{tripathi11core} to remove the diffuse component from their core measurements is reasonable. They estimated a potential error in the EM slope of less than 0.1.}

We caution that the observing geometries are completely different for these two studies and that a horizontal integration through a point above the limb may be much different from vertical integration through the same point. In addition, we could not study the effect of gravitational stratification at the height of the AR core, even by assuming the diffuse emission to be unresolved loop structures. This is because, measuring the loop length is not possible in our case taking the diffuse nature of the region into account. Hence our results may not be comparable with the results of \cite{tripathi11core}. 

Our earlier on-disk study of \textit{AR 10961} \citep{tripathi11core} benefited from a full spectral scan, which we do not have for our present limb study of this region. The lack of spectral lines hotter than $\log\, T = 6.45$ prevents us from fully characterizing the core plasma, but we can obtain a rather complete picture of the diffuse and boundary region plasma. Similar to the analysis performed for \textit{AR 10939}, we have chosen different regions (shown in the top right panel Fig.~\ref{fig4}) and have grouped them in different sets, namely set1, set2 and set3. The unrevised (row 2) and the revised (row 3) EMDs for set~2 and set~3 are plotted in the middle and right panels, and those for set~1 are plotted in the left panels in Fig.~\ref{fig4}. Similar to active region AR 10939, we have plotted the EM for m9 and m10 with pluses and regions m11 and m12 with triangles to differentiate between with regions with and without possible moss contribution. The EMDs for set2 and set3 appear very similar to 
those for \textit{AR 10939}, except that we do not find the peak above $\log\, T = 6.45$. This is because we do not have spectral lines at those temperatures (above \ion{Fe}{16} temperature) in this study. (\cite{tripathi11core} found that the core EM peaked at $\log\, T = 6.55$ when observed on the disk.) Therefore, these data are not adequate for studying the active region core heating and also to probe the variation of slopes of AR core with background subtractions using an averaged diffuse region EMD, as done with the \textit{AR 10939}. However, this region is still useful for studying the diffuse region due to the fact that the EM for diffuse region peaks at $\log\, T = 6.25$ as established earlier for \textit{AR 10939} using the full spectral scan (assuming that the active regions are similar).

The bottom left plot in Fig.~\ref{fig4} shows the EM distribution for diffuse regions named as set1, comprising m1, m2, m3, m4 \& m5. The peak of EMD is again at \ion{Fe}{13} at $\log\, T = 6.25$. We determined the slope of EM curves (between $\log\, T = 6.25$ and $\log\, T = 6.05$) for all these five regions in set1 and studied the ratio of hot (\ion{Fe}{13}) to warm (\ion{Fe}{10} as well as
\ion{Si}{7}) emission. The values are provided in Table~\ref{table_slope_EMratio}. Fig.~\ref{fig4} shows that the EM curves of set1 (high altitude diffuse emission) differ from the EM curves of set3 (core) of \textit{AR~10961} by $\delta$log EM $\approx$ 0.3-0.6 for $\log\, T <~$6.2, which corresponds to roughly a factor of 2-3. Similarly, EM curves of set1 differ from set3 of  \textit{AR~10939} by a factor of 2-4. The EM distribution itself is comparable or slightly larger for \textit{AR~10939} (Fig.~\ref{fig3}) when compared with \textit{AR~10961}.

\begin{table*}[t!]
\small
\begin{center}
\caption{Estimated  power law slope ($\alpha$) and the ratio of hot to warm plasma (\ion{Fe}{10} \& \ion{Si}{7}) of the diffuse emission regions (set1) shown in Fig. \ref{fig3} (top) and  Fig. \ref{fig4} (top), considering both the unrevised and the revised intensity. }
\vspace{0.11in}
\label{table_slope_EMratio}
\begin{tabular} {ccccccccccc}
\hline
& & \multicolumn{3}{l}{\textit{Unrevised}} &  & \multicolumn{3}{l}{\textit{revised}}\citep{giulio13} \\
\hline
& Masked & slope& \ion{Fe}{13}/& \ion{Fe}{13}/ && slope& \ion{Fe}{13}/& \ion{Fe}{13}/ \\
& regions& (${\alpha}$)	& \ion{Fe}{10} & \ion{Si}{7}   &  & (${\alpha}$) &\ion{Fe}{10} & \ion{Si}{7}	\\
\hline
\textit{AR 10939} & 	m1	& 2.90 	&  3.1 & 19.7 &  & 2.04 &  2.3 & 17.2 	\\
		  & 	m2	& 3.08 	&  3.2 & 21.6 &  & 2.22 &  2.4 & 18.8 	\\
		  &	m3	& 2.95 	&  3.3 & 9.5  &  & 2.08 &  2.5 & 8.3  	\\
		  &	m4	& 2.87 	&  3.2 & 14.1 &  & 2.01 &  2.4 & 12.3 	\\	
\hline
\textit{AR 10961}& 	m1	& 2.31	&  2.5 & 15.4 & & 1.44 	&  1.8 & 12.5 	\\	
		  & 	m2	& 2.79	&  2.3 & 37.9 & & 1.92 	&  2.4 & 30.8 	\\	
		  &	m3	& 3.45	&  4.4 & 55.5 & & 2.58 	&  3.2 & 45.1 	\\		
		  & 	m4	& 2.31	&  2.6 & 13.0 & & 1.45 	&  1.9 & 10.6 	\\		
		  & 	m5	& 4.15	&  6.2 & 108.2& & 3.29 	&  4.6 & 87.9 	\\		
	
\hline
\end{tabular}
\end{center}
\end{table*}

\section{Summary and Discussion} \label{conc}

It has now been established that there is a significant amount of plasma at higher coronal temperatures which does not appear in any structured form, such as distinct loops, when viewed with current instrumentation. Rather, the plasma appears as diffuse emission spreading over a large area. The emission from coronal loops is only about 20-30\% higher than this fuzzy emission \citep[see e.g.,][]{giulio03,viall11}. Therefore, just explaining the heating of structured loops is not enough for understanding the heating of solar active regions in particular and the corona in general. This study aims at understanding the characteristics of these diffuse regions and their possible heating scenario. Keeping this science goal in mind, we have probed two active regions, \textit{AR~10939} and \textit{AR~10961}, observed observed off limb on 2007~Jan~26 and 2007~July ~08, respectively, using emission measure diagnostics techniques.

We have distinguished the diffuse emission region from the rest of the AR by tracking the AR from the disk center to the off-limb, which provided us with a better constraint on selecting topologically different areas within the AR. We have chosen different sets of masked regions, designated 1, 2 $\&$ 3, to isolate, respectively, diffuse emission regions, the boundary between the diffuse regions and the core, and the core itself. EMDs have been obtained for each of the masked regions using the Pottasch method  \citep[see e.g.,][]{pottasch63, jordan71, tripathi10moss}. We also compare the EMD of such topologically different regions.

\begin{table*}[t!]
\small
\begin{center}
\caption{Average power law slopes obtained over different sets of masked regions of \textit{AR 10939} }
\vspace{0.11in}
\label{table_avgslope}
\begin{tabular} {ccccc}
\hline
Masked 	regions  & \multicolumn{2}{c}{Power law slope (${\alpha}$) } \\
\hline
	& \multicolumn{2}{c}{\textit{Unrevised}} & \multicolumn{2}{c}{\textit{revised}} \\
\hline
 		& no BG &BG subtracted 	& no BG &BG subtracted	\\
\hline
	set1	& 2.95 &  -		& 2.08	& -		\\
	set2	& 2.97 &  -		& 2.42	& -		\\
	set3a	& 2.42 & {3.24}	& 2.26	& {3.25}	\\
	set3b	& 3.52 & {4.29}	& 2.84	& {3.55} 	\\
\hline
\end{tabular}
\end{center}
\end{table*}

In EMD, the slope $\alpha$ and the emission measure ratio of hot and warm plasma may indicate the possible mechanisms responsible for heating the plasma. For example, a shallower slope suggests that along a given line of sight, there are comparable amounts of hot  ($\approx$ few MK) and warm ($\approx$ 1 MK) plasma co-existing, implying a low frequency nanoflare heating, often called simply impulsive heating.
In comparison, a steeper slope indicates much less warm plasma than hot plasma, implying a high frequency nanoflare heating, often referred to as steady heating, where the plasma is maintained at a given temperature. Much importance has been given to the estimation of the slope $\alpha$ for active region cores and has always been debated between low and high frequency nanoflare heating \citep{tripathi10core, tripathi11core,
wine11, warren11, warren12, viall12,dadashi12,bradshaw12,reep13,wine13}. While the diffuse emission associated with the AR has been poorly understood.

\cite{guennou13} showed that the uncertainty in the estimation of slope can be $\pm$ 1 and depends on both the atomic physics uncertainties and the number of spectral lines used to constrain the distribution. This may represent the upper limit of the errors involved. \cite{tripathi11core} suggested the total uncertainty  involved in the EM estimation to be $\approx$ 50 \%, which includes the radiometric and line fitting errors associated with intensities, and the uncertainties associated with the atomic physics and the Pottasch method.

The estimated power law slope $\alpha$ and the emission measure ratios of all the diffuse emission regions studied here are given in the Table~\ref{table_slope_EMratio}. The EMD of all these regions show that the emission measure peaks at $\log\, T = 6.25$. The distributions show a monotonic increase from $\log\, T = 5.6$ to $\log\, T = 6.25$ and then starts to decrease at higher temperatures. In general, 8 out of the 9 diffuse emission regions analysed here showed a power law slope in between 2.3 $\&$ 3.5, except for the region m5 from \textit{AR 10961}, which shows a much higher slope ($\alpha$=4.18) and EM ratios (108 when taken with \ion{Si}{7} and 6.2 when taken with \ion{Fe}{10}) than the others. Implication of the revised radiometric calibration vastly changes the earlier results. It produced comparatively shallower slopes in the range of 1.4 \& 2.6  and {lowers the hot to warm plasma ratio.}

The EMD of the core regions, which can only be studied for the \textit{AR 10939}, shows consistency with previous studies. The peak of the EM is found to be around $\log\, T=6.55$ which is also consistent with the EM peak obtained in previous studies \citep{tripathi10core, tripathi11core, wine11, warren11, warren12, wine13}.

A sample of masked regions chosen from both the active regions, that are discussed in Tables~\ref{eis_linelist_AR10939_1} $\&$ \ref{eis_linelist_AR10961_1}, can be used to probe the variation of spectroscopic parameters like intensity, density etc., along with the EMD slope ($\alpha$) as a function of distance from the AR core which is just over the limb. These samples comprise of m2 (set1), m5 (set2), m8 $\&$ m10 (set3) for the \textit{AR 10939}  and  m3 (set1), m7 (set2), m9 $\&$ m11 (set3) for the \textit{AR 10961}, that fall as a strip which includes both the core plasma and the diffuse emission plasma above the core. Tables~\ref{eis_linelist_AR10939_1} $\&$  \ref{eis_linelist_AR10961_1} clearly show that the intensities and densities decrease from the core towards the diffuse regions, as a function of distance from the limb, in agreement with \cite{odwyer11}. The power law slope of those sample masked regions from the \textit{AR 10939} are 3.0 (m2), 3.2 (m5), 2.4 (m8) $\&$ 2.7 (m10). Similarly, the
slope of sample masked regions from the \textit{AR 10961} are 3.4 (m3), 3.5 (m7), 3.4 (m9) $\&$ 4.0 (m11). We, however, note that the slopes determined in the core of the active region for AR 10961 does not represent the true EMD slope as a full spectrum for this AR was not available. The highestest temperature line in the available raster is \ion{Fe}{16}, whereas the core emission peaks higher at $\log\, T = 6.45$. Note also that these slopes were measured with the uncorrected data. Slopes measured with the revised calibration are smaller by about 0.9, at least for the diffuse regions where we have studied the difference (Table~\ref{table_slope_EMratio}). Based on the uncorrected data, we conclude that the slope does not have an obvious dependence on distance from solar surface. Table~\ref{table_avgslope} shows the average slopes of EMDs, with and without background/foreground subtraction, for the \textit{AR 10939}. A background/foreground subtraction from the unrevised and revised core EMD using, an 
averaged unrevised and revised EMD of the studied diffuse regions respectively, steepened the slope by {$\approx$ 0.8 $\&$ 0.85}. We stress that the background/foreground subtraction is uncertain due to the unknown variation of diffuse emission with altitude (both its horizontal line-of-sight depth and its emissivity).

The difference in the slopes between the two ARs can be attributed to the fact that they are two different ARs at different instances in their lifetime. At the time of observations, \textit{AR 10939} and \textit{AR 10961} are 6 days and 13 days old (Solar \monitor). The heating properties of active regions (low versus high frequency nanoflares) have been shown to depend on the age of the AR \citep{sp2012,ignacio12} and the magnetic field strength of the AR \citep{warren12,giulio14}. Significant errors can come from the difference in the number of spectral lines constraining the distribution \citep{guennou13}.

The slopes of the EMD curves and the hot to warm ratios for diffuse regions obtained in this study is within the range of values obtained for active regions cores. These results suggest that the diffuse emission regions are heated and maintained in similar way as the hot emission in the core of active regions. The slope is similar among the different parts of the active regions (Table~\ref{table_avgslope}). The EM slopes of the diffuse emission outside the core are generally in the range of 2. Thus, they are consistent with both low frequency and high frequency heating, especially considering the uncertainties in the measurements. However, more detailed analysis with many active regions and theoretical modelling should be performed before making any definite conclusions. The results obtained in this study provide further constraints on the properties of diffuse emission in active regions.

\acknowledgments{}
Hinode is a Japanese mission developed and launched by ISAS/JAXA, collaborating with NAOJ as a domestic partner, NASA and STFC (UK) as international partners. Sci- entific operation of the Hinode mission is conducted by the Hinode science team organized at ISAS/JAXA. This team mainly consists of scientists from institutes in the partner countries. Support for the post-launch operation is provided by JAXA and NAOJ (Japan), STFC (U.K.), NASA, ESA, and NSC (Norway). CHIANTI is a collaborative project involving re- searchers at NRL (USA) RAL (UK), and the Universities of Cambridge (UK), George Mason (USA), and Florence (Italy). The work of JAK was funded by the NASA Supporting Research Program. H.E.M. acknowledges S.T.F.C. We acknowledge useful discussions at the ISSI on Active Region Heating. We thank Dr. Peter Young for his valuable comments and discussions. We thank the anonymous referee for thoughtful comments and a careful reading.


\appendix 
\section{Appendix material}

\begin{table} [!h]
\small
\caption{ List of spectral lines used to study the EM over topologically different areas within active region 10961,
along with the calculated intensities and the spectral line fitting errors of a the rest of the masked regions in the respective wavelength.Intensity units are in ergs cm$^{−2}$ s$^{−1}$ sr$^{−1}$. True errors will also include radiometric calibration errors of about 22\% of the intensities \citep{lang06}, in addition to the fitting errors metioned in this table.}
\label{eis_linelist_AR10939_2}
\begin{tabular} {lrrrrrrr} \\
\hline
Spectral lines& \multicolumn{7}{c}{Intensities and fitting errors} \\
\hline
Ion  & m1 & m3 & m4 & m6 & m7 & m9 & m11 \\
\hline

\ion{Mg}{6} &  127$\pm$  2 &  291$\pm$  2 &  133$\pm$  2 &  312$\pm$  3 &  169$\pm$  2 &  537$\pm$  3 &  407$\pm$  4 \\
\ion{Si}{7} &  111$\pm$  1 &  229$\pm$  2 &  109$\pm$  1 &  241$\pm$  2 &  125$\pm$  1 &  383$\pm$  2 &  279$\pm$  2 \\
\ion{Fe}{8} &  274$\pm$  3 &  522$\pm$  4 &  304$\pm$  4 &  707$\pm$  6 &  552$\pm$  6 & 1267$\pm$  8 & 1253$\pm$ 11 \\
\ion{Fe}{8} &  182$\pm$  2 &  362$\pm$  3 &  196$\pm$  3 &  451$\pm$  4 &  276$\pm$  4 &  789$\pm$  5 &  665$\pm$  7 \\
\ion{Fe}{9} &  139$\pm$  1 &  159$\pm$  1 &   90$\pm$  2 &  157$\pm$  1 &  113$\pm$  1 &  196$\pm$  1 &  167$\pm$  2 \\
\ion{Fe}{9} &  199$\pm$  2 &  244$\pm$  2 &  139$\pm$  1 &  250$\pm$  2 &  172$\pm$  2 &  326$\pm$  3 &  292$\pm$  4 \\
\ion{Fe}{10}& 1937$\pm$ 76 & 1861$\pm$ 71 & 1464$\pm$ 80 & 2076$\pm$ 85 & 2136$\pm$ 99 & 2726$\pm$102 & 3043$\pm$130 \\
\ion{Fe}{10}&  761$\pm$  5 &  699$\pm$  5 &  515$\pm$  5 &  809$\pm$  6 &  737$\pm$  6 & 1056$\pm$  7 & 1186$\pm$ 10 \\
\ion{Fe}{11}& 2315$\pm$ 18 & 2139$\pm$ 17 & 1715$\pm$ 18 & 2526$\pm$ 22 & 2433$\pm$ 23 & 3006$\pm$ 25 & 3894$\pm$ 36 \\
\ion{Fe}{11}& 1401$\pm$  5 & 1294$\pm$  4 & 1035$\pm$  5 & 1474$\pm$  7 & 1449$\pm$  7 & 1817$\pm$  8 & 2585$\pm$ 12 \\
\ion{Si}{10}&  838$\pm$  5 &  682$\pm$  4 &  490$\pm$  4 &  855$\pm$  4 &  908$\pm$  5 & 1198$\pm$  5 & 2384$\pm$  9 \\
\ion{Si}{10}&  392$\pm$  3 &  309$\pm$  3 &  235$\pm$  3 &  384$\pm$  3 &  393$\pm$  3 &  522$\pm$  3 &  921$\pm$  5 \\
\ion{Fe}{12}& 1031$\pm$  3 &  925$\pm$  3 &  708$\pm$  3 & 1135$\pm$  3 & 1118$\pm$  3 & 1394$\pm$  4 & 2374$\pm$  6 \\
\ion{Fe}{12}& 2615$\pm$  4 & 2477$\pm$  4 & 1923$\pm$  4 & 2949$\pm$  5 & 2861$\pm$  6 & 3561$\pm$  6 & 5508$\pm$ 10 \\
\ion{Fe}{13}& 2622$\pm$  8 & 2524$\pm$  7 & 1951$\pm$  8 & 2933$\pm$  9 & 2863$\pm$  9 & 3380$\pm$  9 & 4775$\pm$ 14 \\
\ion{Fe}{13}& 1507$\pm$  7 & 1546$\pm$  7 & 1001$\pm$  7 & 2362$\pm$  9 & 2334$\pm$  9 & 3485$\pm$ 11 & 9171$\pm$ 20 \\
\ion{Fe}{14}& 1195$\pm$  4 & 1237$\pm$  4 &  894$\pm$  4 & 1881$\pm$  5 & 2069$\pm$  6 & 3031$\pm$  7 & 6578$\pm$ 12 \\
\ion{Fe}{14}& 1449$\pm$  4 & 1529$\pm$  4 & 1131$\pm$  5 & 2225$\pm$  5 & 2342$\pm$  6 & 3219$\pm$  6 & 5863$\pm$ 11 \\
\ion{Fe}{15}& 6262$\pm$ 14 & 6346$\pm$ 14 & 5223$\pm$ 15 &11958$\pm$ 19 &12059$\pm$ 21 &18695$\pm$ 25 &24015$\pm$ 36 \\
\ion{S}{13} &  330$\pm$  4 &  294$\pm$  3 &  275$\pm$  4 &  600$\pm$  5 &  987$\pm$  6 & 1234$\pm$  7 & 2095$\pm$ 11 \\
\ion{Fe}{16}&  277$\pm$  2 &  301$\pm$  2 &  287$\pm$  2 &  825$\pm$  4 &  959$\pm$  4 & 1814$\pm$  5 & 2380$\pm$  7 \\
\ion{Ca}{14}&   15$\pm$  1 &   13$\pm$  0 &   15$\pm$  1 &   81$\pm$  1 &   73$\pm$  1 &  206$\pm$  2 &  203$\pm$  2 \\
\ion{Ca}{15}&   83$\pm$ 10 &   93$\pm$ 10 &   82$\pm$ 13 &   52$\pm$  2 &   35$\pm$  2 &  104$\pm$  3 &  114$\pm$  4 \\

\hline
\end{tabular}
\end{table}

\begin{table} [!ht]
\small
\caption{ List of spectral lines used to study the EM over topologically different areas within active region 10961,
along with the calculated intensities and the spectral line fitting errors of a the rest of the masked regions in the respective wavelength.Intensity units are in ergs cm$^{−2}$ s$^{−1}$ sr$^{−1}$. True errors will also include radiometric calibration errors of about 22\% of the intensities \citep{lang06}, in addition to the fitting errors metioned in this table. }
\label{eis_linelist_AR10961_2}
\begin{tabular} {lrrrrrrrr} \\ 
\hline
 & \multicolumn{8}{c}{Intensities and fitting errors} \\
\hline
Ion  & m1 & m2 & m4 & m5 & m6 & m8 & m10 & m12 \\
\hline

\ion{Si}{7}  &   62$\pm$  1 &   25$\pm$  1 &   85$\pm$  1 &   11$\pm$  1 &   121$\pm$  2 &   50$\pm$  1 &  126$\pm$  2 &  153$\pm$  3 \\
\ion{Fe}{8}  &  121$\pm$  3 &   49$\pm$  2 &  158$\pm$  3 &   29$\pm$  3 &   258$\pm$  5 &  141$\pm$  4 &  333$\pm$  5 &  543$\pm$ 10 \\
\ion{Fe}{10} & 1292$\pm$123 & 1139$\pm$ 94 & 1568$\pm$102 &  817$\pm$106 &  1769$\pm$129 & 1380$\pm$118 & 2036$\pm$118 & 2535$\pm$200 \\
\ion{Fe}{10} &  434$\pm$  5 &  324$\pm$  4 &  481$\pm$  4 &  219$\pm$  3 &   580$\pm$  6 &  446$\pm$  5 &  691$\pm$  6 & 1016$\pm$ 10 \\
\ion{Fe}{11} &  767$\pm$  5 &  709$\pm$  4 &  924$\pm$  4 &  689$\pm$  4 &   974$\pm$  5 &  978$\pm$  5 & 1517$\pm$  6 & 2097$\pm$ 12 \\
\ion{Fe}{12} & 1389$\pm$  3 & 1397$\pm$  3 & 1619$\pm$  3 & 1599$\pm$  3 &  1865$\pm$  4 & 1921$\pm$  4 & 2793$\pm$  5 & 3681$\pm$  9 \\
\ion{Fe}{13} & 1317$\pm$  7 & 1351$\pm$  6 & 1552$\pm$  6 & 1724$\pm$  7 &  1819$\pm$  9 & 1945$\pm$  8 & 2662$\pm$  9 & 3117$\pm$ 15 \\
\ion{Fe}{13} &  541$\pm$  6 &  502$\pm$  4 &  610$\pm$  5 &  655$\pm$  5 &   985$\pm$  8 & 1008$\pm$  7 & 2229$\pm$  9 & 4356$\pm$ 20 \\
\ion{Fe}{14} &  430$\pm$  3 &  390$\pm$  2 &  456$\pm$  2 &  495$\pm$  3 &   785$\pm$  4 &  789$\pm$  3 & 1623$\pm$  5 & 3112$\pm$ 10 \\
\ion{Fe}{14} &  528$\pm$  4 &  499$\pm$  3 &  591$\pm$  4 &  621$\pm$  4 &   917$\pm$  6 &  948$\pm$  5 & 1631$\pm$  6 & 2614$\pm$ 11 \\
\ion{Fe}{15} & 1543$\pm$  7 & 1327$\pm$  6 & 1434$\pm$  6 & 1447$\pm$  7 &  3300$\pm$ 12 & 3193$\pm$ 10 & 5219$\pm$13  & 8706$\pm$ 26 \\
\ion{Fe}{16} &   57$\pm$  3 &   41$\pm$  2 &   52$\pm$  2 &   41$\pm$  2  &  224$\pm$  4 &  177$\pm$  3 & 402$\pm$  4  & 723$\pm$  7  \\

\hline
\end{tabular}
\end{table}

\end{document}